\documentclass{nature}
\usepackage[english]{babel}
\usepackage{url}
\usepackage[utf8x]{inputenc}
\usepackage{amsmath}
\usepackage{graphicx}
\usepackage[T1]{fontenc}
\usepackage{xcolor}
\usepackage{soul}
\usepackage{aas_macros}
\usepackage{hyperref}
\newcommand\Colorhref[3][gray]{\href{#2}{\small\color{#1}#3}}

\title{Machine Learning and the future of \\ Supernova Cosmology}

\author{Emille E. O. Ishida$^1$}

\begin{document}
\maketitle

\begin{affiliations}
 \item Universit\'{e} Clermont Auvergne, CNRS/IN2P3, LPC, F-63000 Clermont-Ferrand, France
\end{affiliations}

At the end of the 20th century, extragalactic distance measurements based on type Ia supernovae (SNe Ia) provided the first evidence that our Universe is  currently undergoing an accelerated expansion \cite{1998AJ....116.1009R, 1999ApJ...517..565P}.  A result that was subsequently confirmed by a series of independent probes \cite{2018RPPh...81a6901H}, each contributing with different pieces of  what is known as the standard model of cosmology. Nevertheless, two decades into the 21st century, a fundamental theory concerning the physics of dark energy -- the energy component causing the cosmic acceleration -- is still missing. In a remarkable community effort, astronomers have devoted a large fraction of their resources to imposing more restrictive constraints over cosmological parameters -- in the hope that they might shed some light on the properties of dark energy. In this new scenario, SNe Ia continue to play a central role as cosmological standardizable candles -- and consequently, feature among the main targets of  modern large-scale sky surveys. 

This community interest in a specific type of SN for cosmological applications, coupled with rapid technological developments, bigger telescopes and the advent of surveys systematically monitoring large portions of the sky, raised another set of challenges. Among them, the overwhelming task of accurately classifying an unprecedented large number of SN candidates given very limited spectroscopic resources. 

After a transient is identified as a SN candidate, its true class must be determined via spectroscopy (SN classification is based on the presence/absence of specific spectral features\cite{Gal-Yam2017}). In case a SN Ia is confirmed, its photometric light curve and redshift can be employed in the cosmological analysis. This is the first bottleneck in the SN cosmology pipeline. Since spectroscopy will always be a scarce -- and very expensive --  resource, the majority of the photometrically identified SN candidates will never be spectroscopically confirmed.  Although redshift information can be obtained from the host galaxy,  if we aim to fully exploit the cosmological potential of the transient sky it is mandatory to develop strategies for inferring SN classification based purely on photometric data.  For SN cosmology, the big data paradigm is approaching fast. 

The first reports on dark energy made use of less than 100 SNe, while current spectroscopic samples used for cosmology encompass less than 2,000 objects. Meanwhile, the Dark Energy Survey\footnote{\Colorhref{https://www.darkenergysurvey.org/}{https://www.darkenergysurvey.org/}} (DES) detected  ~12,000 SN candidates in its first 3 observational seasons \cite{2018arXiv181109565D} and the current ongoing Zwicky Transient Facility\footnote{\Colorhref{https://www.ztf.caltech.edu/}{https://www.ztf.caltech.edu/}} (ZTF) is expected to detect 600 SNe Ia light curves per year in low redshift only ($z_{med} \sim$ 0.1) \cite{2019arXiv190203923F}. This situation will get even more drastic with the advent of  The Large Synoptic Survey Telescope\footnote{\Colorhref{https://www.lsst.org}{https://www.lsst.org}} (LSST), scheduled to begin operations in 2021, which will measure ~300,000 well-sampled SNe Ia light curves over 10 years \cite{2018arXiv181200515L}, with less than 3\% expected to be spectroscopically confirmed \cite{2019Msngr.175...58S}. At the same time, at least a few thousand SN candidates sit in databases from past and current observational campaigns, waiting for proper classification. Efforts towards employing these light curves in purely photometric SN cosmology are underway \cite{2018ApJ...857...51J}, but they assume the existence of a reliable classifier.

The photometric SN classification task: mapping light curve shapes into classes which were initially defined according to the presence/absence of spectroscopic features, can be described as a supervised machine learning (ML) problem. In this scenario, the small set of light curves spectroscopically  classified (hereafter, the training sample) is used to train ML algorithms which will subsequently be applied to the non-labelled light curves (hereafter, the target sample). Moreover, given the large volume of data to be delivered by current and future surveys, we are left with few alternatives beyond developing automated classification tools. Nevertheless, the translation of text-book ML algorithms to the astronomical scenario is far from being straightforward.  The first obstacle to be overcome is the absence of a  training sample that is statistically representative of the target sample. Due to the strong requirements demanded by spectroscopic observations -- good observation conditions and significantly large integration times -- and the increasing interest in  cosmological applications, our current training data is strongly biased towards high signal to noise ratio (SNR) objects, and over-populated by SNe Ia. This is a problem for ML models that learn by example and are not expected to extrapolate their capabilities beyond cases present in training data.

This problem has been investigated by the engaged community for a long time. The first systematic attempt in quantifying the potential and limitations of ML applications to the SN classification problem was the SuperNova Photometric Classification Challenge (SNPCC). It consisted of a synthetic data set of ~20,000 SN light curves simulated according to the specifications for the DES. Among these SNe, labels were provided for ~1,100 objects, chosen to mimic the biases present in a real spectroscopically classified sample. Researchers were invited to apply their algorithms to this data set and a total of 10 different groups submitted their answers \cite{2010PASP..122.1415K}. The SNPCC was a milestone in the community efforts towards the development of automated SN photometric classification algorithms. Although none of the entries performed significantly better than all the others, the resulting data set became the standard testing ground for new ideas in the following decade \cite{Lochner2016}.  

The availability of a realistic simulated data set allowed different groups to test a wide set of algorithms \cite{2013MNRAS.430..509I, 2018MNRAS.477.4142D}, compare results \cite{Lochner2016}, and sparked the investigation of strategies to overcome known biases. The most popular among these being semi-supervised learning and data augmentation. Semi-supervised learning uses light curve shape information present in the target data in a preliminary feature extraction phase where both samples, training and target, are used in the determination of a suitable low-dimensional representation. \cite{2012MNRAS.419.1121R} reported limited capabilities of this approach using features from a diffusion map and a random forest classifier. Results from the original SNPCC training sample were not satisfactory (50\% purity) but improved significantly when using a deeper magnitude-limited (more representative) training sample (72\% purity).  Data augmentation comprises a large set of strategies whose goal is to transform/increment the training data in the hope it  becomes a better representation of the target sample. Revsbech \textit{et al.} \cite{10.1093/mnras/stx2570} explored this strategy by drawing additional light curves from Gaussian process fit to instances of underrepresented classes.

The positive impact of the SNPCC data encouraged the development of a second challenge. The Photometric LSST Astronomical Time-series Classification Challenge\footnote{\Colorhref{https://www.kaggle.com/c/PLAsTiCC-2018}{https://www.kaggle.com/c/PLAsTiCC-2018}} (PLAsTiCC) was hosted by the Kaggle platform from September to December 2018 and built to mimic 3 years worth of LSST observations. The training sample contained ~8,000 objects from 14 different classes while the target sample held ~3 million light curves from 14+1 classes. The extra class in the target sample encapsulated 5 different transient models of rare or yet-to-be-confirmed sources expected to be observed by LSST. The challenge attracted more than 1,000 participants, with the best-scoring entry \cite{2019arXiv190704690B} relying heavily on data augmentation and using a tree-based algorithm. The complete repercussions of PLAsTiCC data and the scientific impact from the many strategies proposed to address it are still to be quantified. Nevertheless, as happened before with the SNPCC, it  provided a fertile ground for the development of simulation and analysis tools and resulted in a realistic synthetic data set \cite{plasticc} that will play a crucial role in preparing for the arrival of LSST data.

Parallel to all these efforts, a few groups investigated  the applicability of neural networks \cite{2013MNRAS.429.1278K} and deep learning (DL)  \cite{2017ApJ...837L..28C} algorithms to this problem. In their original form, these techniques require a significantly larger  training sample when compared to traditional ML strategies. For astronomy, however, the absence of such training meant that algorithms must be adapted. In an insightful example, Pasquet \textit{et al.}\cite{2019A&A...627A..21P} recently presented PELICAN (deeP architecturE for the LIght Curve ANalysis), a framework that combines data augmentation, semi-supervised learning and DL, via convolutional neural networks, as a strategy to overcome the non-representativeness issue.  The first module of PELICAN employs the entire data set to learn a low dimensional representation of the data using autoencoders (feature extraction stage). In this new parameter space, the second module uses the available labels to identify pairs of fainter/brighter objects of the same class and optimizes a loss function which decreases the distances within such pairs, translating the effect of redshift,  before feeding them to the classification module. This approach showed promising results when applied to the SNPCC data. Comparison with results reported by Richards \textit{et al.}\cite{2012MNRAS.419.1121R}, makes it clear that the complexity of PELICAN does translate into a higher efficiency in grasping more layers of information from the same data set. The authors also reported results from a transfer learning task -- where the framework was trained with simulations for the Sloan Digital Sky Survey\footnote{\Colorhref{https://www.sdss.org/}{https://www.sdss.org/}} and subsequently used to classify real, spectroscopically confirmed, data from the same telescope. Although results from training exclusively on simulations were poor, the inclusion of a small number of real light curves in the training stage highly improved the performance of the classifier -- reflecting that there are still important characteristics of the real data to be added to the simulations.

All the examples described so far tackle the classification of entire light curves and aim to optimize the use of purely photometric data once decisions regarding spectroscopy were  already made. There is however, another possibility to approach the problem: if we had a photometric classifier able to deliver reliable results based on partial light curves (e.g. considering only epochs up to around maximum brightness) this information could be used to guide the allocation of spectroscopic follow-up resources. How to do this is a crucial question already posed in the SNPCC, which received no answers at the time.

Fortunately, progress in simulations as well as DL techniques allowed interesting developments in this direction.  Recently, M\"oller and Boissi\`ere\cite{2019arXiv190106384M} presented SUPERNNOVA, a framework that uses a bidirectional recurrent neural network (RNN) in the development of a classifier able to deliver time-dependent class probabilities. Moreover, by using a Bayesian adaptation of the RNN architecture, the framework returns a posterior distribution for each classification, allowing a detailed calibration study of resulting probabilities. The method relies on the availability of a large, balanced and full light curve training sample which can only be achieved through simulations -- the application to real data  then constituting a transfer learning approach. Reported results, based on DES simulations, indicate that the method performs at its best for binary ‘Ia or not Ia’ classifications estimated around maximum brightness. The Bayesian adaptation also allows   the possibility to identify models not presented during training -- or out-of-distribution events. The authors report noticeably larger posterior distributions when attempting to classify extreme cases of non-realistic light curves.

Partial light curve analysis can be pushed a little further if we consider the need for very early classifications -- due to the necessary time to coordinate spectroscopic follow-up as well as for dealing with fast-evolving non-SN-like transients. This scenario was tackled by Muthukrishna \textit{et al.}\cite{2019arXiv190400014M}, who proposed RAPID (Real-time Automated Photometric IDentification)  -- a unidirectional RNN approach aimed to provide probabilistic classifications from 1 day after the initial detection to the final epoch of the complete light curve observation. The framework was tested in simulations for the Zwicky Transient Facility (ZTF) using 12 different transient models from PLAsTiCC and provided encouraging results even from as early as 2 days from the trigger of SNe Ia, superluminous supernovae (SLSNe) and Intermediate Luminosity Transients (ILOTs).  Once more, the application of a DL strategy to real data translates into a transfer learning approach. The authors report successful outcomes from applying RAPID to a few real light curves from ZTF as a demonstration of the applicability of the method. 

Results from early classification frameworks like SUPERNNOVA or RAPID will be extremely valuable in the new data paradigm imposed by LSST -- whose difference image analysis pipeline (DIA) is expected to produce up to 10 million alerts per night. This alert stream will be distributed to a few community brokers whose task is to filter, classify and redistribute enhanced alert information to the larger astronomical community. In such a big data context,   early classification algorithms which can guide decisions on spectroscopic follow-up allocation will be paramount. Nevertheless, we are still missing an automated protocol under which such decisions are taken. If, on the one hand,  following targets with a high probability of being a SN Ia would increase the number of available samples for cosmology,  on the other it would also result in highly biased training samples, making it impossible to use  light curves without spectroscopic confirmation. If we ever aim to take advantage of the larger fraction of light curves measured by LSST it is imperative to devise strategies that would lead to more representative training samples -- and consequently improve the performance of any ML classifier. 

The COsmostatistics INitiative\footnote{\Colorhref{https://cosmostatistics-initiative.org/}{https://cosmostatistics-initiative.org/}} (COIN) proposed the use of active learning (AL) techniques in approaching this problem \cite{2019MNRAS.483....2I}. AL is a class of ML methods developed to optimize the construction of informative training samples in situations where labeling is expensive, risky or time consuming, for example, speech-to-text translations or pharmaceutical trials. In this strategy, a ML model is trained and subsequently used to identify which of the objects in the target sample would be more informative to the model if labeled. Once this instance is identified a query is made (in our case a spectrum is requested) and the new labelled light curve is added to the training sample, allowing the entire model to be updated. Results based on the SNPCC data show promising results even in the absence of an initial training sample. Starting from a random classifier the algorithm avoids the need to remove biases from sub-optimal training and concentrates the allocation of spectroscopic time in highly informative objects.  Moreover, AL techniques can be an interesting addition to community brokers  with strong ML components  - e.g., ANTARES\footnote{\Colorhref{https://antares.noao.edu}{https://antares.noao.edu}} \cite{2018ApJS..236....9N}, ALeRCE\footnote{\Colorhref{http://alerce.science/}{http://alerce.science/}} and Fink\footnote{\Colorhref{https://fink-broker.readthedocs.io/en/latest/}{https://fink-broker.readthedocs.io/en/latest/}}. Training samples constructed following such strategies would enable an optimal exploitation of the large target sample for cosmological as well as astrophysical applications. 

The development of a complete pipeline allowing optimal SN photometric classification is a difficult challenge to be overcome before we can exploit the full potential of large-scale transient surveys for SN cosmology. A final answer to this question will probably not come from a single group or strategy, it is more likely that a combination of efforts will result in a yet-to-be-known framework comprising elements from many different sources. Once this is achieved however, it will be important to keep in mind that this is still only part of the difficult path towards fully photometric SN cosmology. The final pipeline will also need to take into account the presence of  probabilistic classifications in cosmological parameter estimation \cite{2018ApJ...857...51J} and the proper determination of distance bias \cite{2019MNRAS.485.1171K} given different strategies for selection of spectroscopic samples. There is certainly a long way to go but the recent developments within the astronomical community, as well as the high level of engagement of scientists from other research fields, show there is a huge potential for development on multiple fronts. 

There are hidden aspects of our Universe waiting to be discovered in the data deluge that will hit us soon -- and everything indicates we are in the correct path to find them.

\section*{Acknowledgments}

EEOI thanks Rafael S. de Souza and Alberto Krone-Martins for comments on this manuscript. EEOI is supported by a 2018 CNRS MOMENTUM fellowship.

\bibliography{ref}

\begin{thebibliography}{10}
\expandafter\ifx\csname url\endcsname\relax
  \def\url#1{\texttt{#1}}\fi
\expandafter\ifx\csname urlprefix\endcsname\relax\def\urlprefix{URL }\fi
\providecommand{\bibinfo}[2]{#2}
\providecommand{\eprint}[2][]{\url{#2}}

\bibitem{1998AJ....116.1009R}
\bibinfo{author}{{Riess}, A.~G.} \emph{et~al.}
\newblock \bibinfo{title}{{Observational Evidence from Supernovae for an
  Accelerating Universe and a Cosmological Constant}}.
\newblock \emph{\bibinfo{journal}{\aj}} \textbf{\bibinfo{volume}{116}},
  \bibinfo{pages}{1009--1038} (\bibinfo{year}{1998}).
\newblock \eprint{astro-ph/9805201}.

\bibitem{1999ApJ...517..565P}
\bibinfo{author}{{Perlmutter}, S.} \emph{et~al.}
\newblock \bibinfo{title}{{Measurements of {\ensuremath{\Omega}} and
  {\ensuremath{\Lambda}} from 42 High-Redshift Supernovae}}.
\newblock \emph{\bibinfo{journal}{\apj}} \textbf{\bibinfo{volume}{517}},
  \bibinfo{pages}{565--586} (\bibinfo{year}{1999}).
\newblock \eprint{astro-ph/9812133}.

\bibitem{2018RPPh...81a6901H}
\bibinfo{author}{{Huterer}, D.} \& \bibinfo{author}{{Shafer}, D.~L.}
\newblock \bibinfo{title}{{Dark energy two decades after: observables, probes,
  consistency tests}}.
\newblock \emph{\bibinfo{journal}{Reports on Progress in Physics}}
  \textbf{\bibinfo{volume}{81}}, \bibinfo{pages}{016901}
  (\bibinfo{year}{2018}).
\newblock \eprint{1709.01091}.

\bibitem{Gal-Yam2017}
\bibinfo{author}{Gal-Yam, A.}
\newblock \emph{\bibinfo{title}{Observational and Physical Classification of
  Supernovae}}, \bibinfo{pages}{1--43} (\bibinfo{publisher}{Springer
  International Publishing}, \bibinfo{address}{Cham}, \bibinfo{year}{2017}).
\newblock \urlprefix\url{https://doi.org/10.1007/978-3-319-20794-0_35-1}.

\bibitem{2018arXiv181109565D}
\bibinfo{author}{{D'Andrea}, C.~B.} \emph{et~al.}
\newblock \bibinfo{title}{First cosmology results using type ia supernovae from
  the dark energy survey: Survey overview and supernova spectroscopy}.
\newblock \emph{\bibinfo{journal}{arXiv e-prints}}
  \bibinfo{pages}{arXiv:1811.09565} (\bibinfo{year}{2018}).
\newblock \eprint{1811.09565}.

\bibitem{2019arXiv190203923F}
\bibinfo{author}{{Feindt}, U.} \emph{et~al.}
\newblock \bibinfo{title}{{$\texttt{simsurvey}$: Estimating Transient Discovery
  Rates for the Zwicky Transient Facility}}.
\newblock \emph{\bibinfo{journal}{arXiv e-prints}}
  \bibinfo{pages}{arXiv:1902.03923} (\bibinfo{year}{2019}).
\newblock \eprint{1902.03923}.

\bibitem{2018arXiv181200515L}
\bibinfo{author}{{Lochner}, M.} \emph{et~al.}
\newblock \bibinfo{title}{Optimizing the lsst observing strategy for dark
  energy science: Desc recommendations for the wide-fast-deep survey}.
\newblock \emph{\bibinfo{journal}{arXiv e-prints}}
  \bibinfo{pages}{arXiv:1812.00515} (\bibinfo{year}{2018}).
\newblock \eprint{1812.00515}.

\bibitem{2019Msngr.175...58S}
\bibinfo{author}{{Swann}, E.} \emph{et~al.}
\newblock \bibinfo{title}{{4MOST Consortium Survey 10: The Time-Domain
  Extragalactic Survey (TiDES)}}.
\newblock \emph{\bibinfo{journal}{The Messenger}}
  \textbf{\bibinfo{volume}{175}}, \bibinfo{pages}{58--61}
  (\bibinfo{year}{2019}).
\newblock \eprint{1903.02476}.

\bibitem{2018ApJ...857...51J}
\bibinfo{author}{{Jones}, D.~O.} \emph{et~al.}
\newblock \bibinfo{title}{{Measuring Dark Energy Properties with
  Photometrically Classified Pan-STARRS Supernovae. II. Cosmological
  Parameters}}.
\newblock \emph{\bibinfo{journal}{\apj}} \textbf{\bibinfo{volume}{857}},
  \bibinfo{pages}{51} (\bibinfo{year}{2018}).
\newblock \eprint{1710.00846}.

\bibitem{2010PASP..122.1415K}
\bibinfo{author}{{Kessler}, R.} \emph{et~al.}
\newblock \bibinfo{title}{{Results from the Supernova Photometric
  Classification Challenge}}.
\newblock \emph{\bibinfo{journal}{\pasp}} \textbf{\bibinfo{volume}{122}},
  \bibinfo{pages}{1415} (\bibinfo{year}{2010}).
\newblock \eprint{1008.1024}.

\bibitem{Lochner2016}
\bibinfo{author}{Lochner, M.}, \bibinfo{author}{McEwen, J.~D.},
  \bibinfo{author}{Peiris, H.~V.}, \bibinfo{author}{Lahav, O.} \&
  \bibinfo{author}{Winter, M.~K.}
\newblock \bibinfo{title}{{PHOTOMETRIC} {SUPERNOVA} {CLASSIFICATION} {WITH}
  {MACHINE} {LEARNING}}.
\newblock \emph{\bibinfo{journal}{The Astrophysical Journal Supplement Series}}
  \textbf{\bibinfo{volume}{225}}, \bibinfo{pages}{31} (\bibinfo{year}{2016}).
\newblock \urlprefix\url{https://doi.org/10.3847%2F0067-0049%2F225%2F2%2F31}.

\bibitem{2013MNRAS.430..509I}
\bibinfo{author}{{Ishida}, E.~E.~O.} \& \bibinfo{author}{{de Souza}, R.~S.}
\newblock \bibinfo{title}{{Kernel PCA for Type Ia supernovae photometric
  classification}}.
\newblock \emph{\bibinfo{journal}{\mnras}} \textbf{\bibinfo{volume}{430}},
  \bibinfo{pages}{509--532} (\bibinfo{year}{2013}).
\newblock \eprint{1201.6676}.

\bibitem{2018MNRAS.477.4142D}
\bibinfo{author}{{Dai}, M.}, \bibinfo{author}{{Kuhlmann}, S.},
  \bibinfo{author}{{Wang}, Y.} \& \bibinfo{author}{{Kovacs}, E.}
\newblock \bibinfo{title}{{Photometric classification and redshift estimation
  of LSST Supernovae}}.
\newblock \emph{\bibinfo{journal}{\mnras}} \textbf{\bibinfo{volume}{477}},
  \bibinfo{pages}{4142--4151} (\bibinfo{year}{2018}).
\newblock \eprint{1701.05689}.

\bibitem{2012MNRAS.419.1121R}
\bibinfo{author}{{Richards}, J.~W.}, \bibinfo{author}{{Homrighausen}, D.},
  \bibinfo{author}{{Freeman}, P.~E.}, \bibinfo{author}{{Schafer}, C.~M.} \&
  \bibinfo{author}{{Poznanski}, D.}
\newblock \bibinfo{title}{{Semi-supervised learning for photometric supernova
  classification}}.
\newblock \emph{\bibinfo{journal}{\mnras}} \textbf{\bibinfo{volume}{419}},
  \bibinfo{pages}{1121--1135} (\bibinfo{year}{2012}).
\newblock \eprint{1103.6034}.

\bibitem{10.1093/mnras/stx2570}
\bibinfo{author}{Revsbech, E.~A.}, \bibinfo{author}{Trotta, R.} \&
  \bibinfo{author}{van Dyk, D.~A.}
\newblock \bibinfo{title}{{STACCATO: a novel solution to supernova photometric
  classification with biased training sets}}.
\newblock \emph{\bibinfo{journal}{Monthly Notices of the Royal Astronomical
  Society}} \textbf{\bibinfo{volume}{473}}, \bibinfo{pages}{3969--3986}
  (\bibinfo{year}{2017}).
\newblock \urlprefix\url{https://doi.org/10.1093/mnras/stx2570}.

\bibitem{2019arXiv190704690B}
\bibinfo{author}{{Boone}, K.}
\newblock \bibinfo{title}{{Avocado: Photometric Classification of Astronomical
  Transients with Gaussian Process Augmentation}}.
\newblock \emph{\bibinfo{journal}{arXiv e-prints}}
  \bibinfo{pages}{arXiv:1907.04690} (\bibinfo{year}{2019}).
\newblock \eprint{1907.04690}.

\bibitem{plasticc}
\bibinfo{author}{Narayan, G.} \& \bibinfo{author}{{et al.}}
\newblock \bibinfo{title}{Unblinded data for plasticc classification challenge}
  (\bibinfo{year}{2019}).
\newblock \urlprefix\url{https://doi.org/10.5281/zenodo.2539456}.

\bibitem{2013MNRAS.429.1278K}
\bibinfo{author}{{Karpenka}, N.~V.}, \bibinfo{author}{{Feroz}, F.} \&
  \bibinfo{author}{{Hobson}, M.~P.}
\newblock \bibinfo{title}{{A simple and robust method for automated photometric
  classification of supernovae using neural networks}}.
\newblock \emph{\bibinfo{journal}{\mnras}} \textbf{\bibinfo{volume}{429}},
  \bibinfo{pages}{1278--1285} (\bibinfo{year}{2013}).
\newblock \eprint{1208.1264}.

\bibitem{2017ApJ...837L..28C}
\bibinfo{author}{{Charnock}, T.} \& \bibinfo{author}{{Moss}, A.}
\newblock \bibinfo{title}{{Deep Recurrent Neural Networks for Supernovae
  Classification}}.
\newblock \emph{\bibinfo{journal}{\apjl}} \textbf{\bibinfo{volume}{837}},
  \bibinfo{pages}{L28} (\bibinfo{year}{2017}).
\newblock \eprint{1606.07442}.

\bibitem{2019A&A...627A..21P}
\bibinfo{author}{{Pasquet}, J.}, \bibinfo{author}{{Pasquet}, J.},
  \bibinfo{author}{{Chaumont}, M.} \& \bibinfo{author}{{Fouchez}, D.}
\newblock \bibinfo{title}{{PELICAN: deeP architecturE for the LIght Curve
  ANalysis}}.
\newblock \emph{\bibinfo{journal}{\aap}} \textbf{\bibinfo{volume}{627}},
  \bibinfo{pages}{A21} (\bibinfo{year}{2019}).
\newblock \eprint{1901.01298}.

\bibitem{2019arXiv190106384M}
\bibinfo{author}{{M{\"o}ller}, A.} \& \bibinfo{author}{{de Boissi{\`e}re}, T.}
\newblock \bibinfo{title}{{SuperNNova: an open-source framework for Bayesian,
  Neural Network based supernova classification}}.
\newblock \emph{\bibinfo{journal}{arXiv e-prints}}
  \bibinfo{pages}{arXiv:1901.06384} (\bibinfo{year}{2019}).
\newblock \eprint{1901.06384}.

\bibitem{2019arXiv190400014M}
\bibinfo{author}{{Muthukrishna}, D.}, \bibinfo{author}{{Narayan}, G.},
  \bibinfo{author}{{Mandel}, K.~S.}, \bibinfo{author}{{Biswas}, R.} \&
  \bibinfo{author}{{Hlo{\v{z}}ek}, R.}
\newblock \bibinfo{title}{{RAPID: Early Classification of Explosive Transients
  using Deep Learning}}.
\newblock \emph{\bibinfo{journal}{arXiv e-prints}}
  \bibinfo{pages}{arXiv:1904.00014} (\bibinfo{year}{2019}).
\newblock \eprint{1904.00014}.

\bibitem{2019MNRAS.483....2I}
\bibinfo{author}{{Ishida}, E.~E.~O.} \emph{et~al.}
\newblock \bibinfo{title}{{Optimizing spectroscopic follow-up strategies for
  supernova photometric classification with active learning}}.
\newblock \emph{\bibinfo{journal}{\mnras}} \textbf{\bibinfo{volume}{483}},
  \bibinfo{pages}{2--18} (\bibinfo{year}{2019}).
\newblock \eprint{1804.03765}.

\bibitem{2018ApJS..236....9N}
\bibinfo{author}{{Narayan}, G.} \emph{et~al.}
\newblock \bibinfo{title}{{Machine-learning-based Brokers for Real-time
  Classification of the LSST Alert Stream}}.
\newblock \emph{\bibinfo{journal}{\apjs}} \textbf{\bibinfo{volume}{236}},
  \bibinfo{pages}{9} (\bibinfo{year}{2018}).
\newblock \eprint{1801.07323}.

\bibitem{2019MNRAS.485.1171K}
\bibinfo{author}{{Kessler}, R.} \emph{et~al.}
\newblock \bibinfo{title}{{First cosmology results using Type Ia supernova from
  the Dark Energy Survey: simulations to correct supernova distance biases}}.
\newblock \emph{\bibinfo{journal}{\mnras}} \textbf{\bibinfo{volume}{485}},
  \bibinfo{pages}{1171--1187} (\bibinfo{year}{2019}).
\newblock \eprint{1811.02379}.

\end{thebibliography}
\end{document}